\documentclass[iop,apjl]{emulateapj}

\def\msun{{\rm ~M}_{\odot}}
\def\rsun{{\rm ~R}_{\odot}}

\usepackage{epstopdf} 
\usepackage{graphicx}
\usepackage{amsmath,amssymb}  

\usepackage{ifpdf}

\ifpdf
\usepackage[pdftex,usenames,dvips]{color}
\else
\usepackage[usenames,dvips]{color}
\fi

\newcommand{\aCE}{\alpha_{\mathrm{CE}}}
\newcommand{\der}[2]{\ensuremath{\frac{d \,#1}{d#2}}}

\begin{document}

\title{Adiabatic Mass Loss and the Outcome of the Common Envelope Phase of Binary Evolution}
\author{Christopher J. Deloye\altaffilmark{1,2}
\email{cjdeloye@gmail.com}
Ronald E. Taam\altaffilmark{1,3}
\email{r-taam@northwestern.edu}}

\altaffiltext{1}{Northwestern University, Department of Physics and Astronomy,
2131 Tech Drive, Evanston, IL 60208}
\altaffiltext{2} {Current address: Mitre Corporation, 7515 Colshire Drive, McLean, VA 22102}
\altaffiltext{3}{Academia Sinica Institute of Astrophysics and Astronomy-TIARA,
P.O. Box 23-141, Taipei, 10617 Taiwan}

\begin{abstract}
We have developed a new method for calculating common envelope (CE) events based on explicit consideration of the 
donor star's structural response to adiabatic mass loss.  In contrast to existing CE prescriptions, 
which specify \textit{a priori} the donor's remnant mass, we determine this quantity self-consistently 
and find it depends on binary and CE parameters. This aspect of our model is particularly important
to realistic modeling for upper main sequence star donors without strongly degenerate cores (and hence without a clear core/envelope boundary). 
We illustrate the central features of our method by considering CE events involving 10 $\msun$ donors on or 
before their red giant branch.  For such donors, the remnant core mass can be as 
much as 30\% larger than the star's He-core mass.  Applied across a population of such binaries, our methodology 
results in a significantly broader remnant mass and final orbital separation distribution and a 20\% increase 
in CE survival rates as compared to previous prescriptions for the CE phase.
\end{abstract}


\section{Introduction}

The common envelope (CE) phase of binary evolution has become a standard paradigm for the formation of short 
period systems containing a compact object since it was introduced by Paczynski (1976) (see reviews by 
Iben \& Livio 1993; Taam \& Sandquist 2000).  While we lack quantitative criteria for determining whether the envelope is successfully ejected in a CE event and a short-period binary system formed, prescriptions have been proposed to approximate 
the relation between the properties of the post-CE binary and its pre-CE progenitor system. The major source of 
uncertainty in these prescriptions is parameterized by a factor, $\aCE$, denoting the efficiency by which 
orbital energy is converted to the kinetic energy of the ejected envelope (see Iben \& Tutukov 1985) or
a factor, $\gamma$, describing the specific angular momentum of the ejected matter (see Nelemans et al. 2000; 
Nelemans \& Tout 2005).  

In recent years, significant effort has been devoted to understanding the early evolution of CE  
events in three dimensions (e.g. Ricker \& Taam 2008) in order to elucidate the relevant hydrodynamic 
processes. However, a detailed description of the final phase, where the envelope is fully ejected and the 
remnant binary system forms, is difficult to obtain via hydrodynamic simulations since resolving the spatial 
scales of the donor's core required prohibitively short time steps.  The main goal of this letter is to describe 
a physically motivated yet computationally tractable methodology for determining the end-state of this final CE 
phase.
  
The outcome of a CE event depends not only on the rate at which mass is ejected, but also on 
the ability of the donor's remnant layers to contract in response to the rapid mass loss.  
The \emph{initial} response of stars to adiabatic mass loss was carried out in the pioneering work of 
Hjellming (1989), where the central focus was on determining when dynamical mass transfer instabilities 
occur.  However the donor's response down to the removal of nearly the entire envelope (relevant to the final 
phase of CE evolution) was not considered.  Here we report on our 
efforts to extend the work of Hjellming (1989) by determining the structural response of donor stars to 
adiabatic mass loss sequences that terminate deep within the donors' core.  In \S 2, we briefly describe these 
computations and their results.  In \S 3, we apply these results to develop a new procedure for determining 
the outcome of the CE phase.  We compare the results of an analysis for a binary population of a $10 \msun$ 
donor with those based on previous prescriptions for the properties of the remnant core in \S 4 and summarize in 
the last section. 
 
\section{Methodology and Donor Response}

The structure and response of a donor star is calculated using a code developed and described in Deloye et 
al.  (2007) in the limit of adiabatic mass loss (i.e., we assume that in the CE, mass loss proceeds at a sufficiently high rate that energy transport between mass layers can be neglected).  The equations governing the donor's structure are given by the condition of 
mechanical equilibrium subject to the constraint that the entropy of a given mass layer is fixed. 
The set of equations used to calculate the donor's structure is equivalent to the following:
\begin{equation}
\der{\ln P}{m} = -\frac{G m}{4 \pi r^4 P}
\label{eq:star_P}
\end{equation}
\begin{equation}
\der{\ln r}{m} = \frac{1}{4 \pi r^3 \rho}
\label{eq:star_r}
\end{equation}
\begin{equation}
\ln s(\rho, T) = \ln s_i(m)
\label{eq:star_s}
\end{equation}
where $\rho$, $T$, $P$, $r$, and $s$ are the density, temperature, pressure, radius, and specific entropy for 
each mesh shell, and $m$ is the mass interior to it. The initial entropy $s_i(m)$ and composition profile, 
$X(m)$, is determined from a donor's initial structure. 

We apply standard boundary conditions except that in lieu of the effective 
temperature-luminosity relation, we enforce the fixed-entropy condition at the exterior boundary.  
To enforce mass loss on the donor's structure, the outer mesh point $m$ equals the varying 
total donor mass, $M_2$.

In reality, the above set of equations is cast in terms of modified $P$, $T$, $m$, and $r$ variables and 
supplemented by a mesh spacing function that automatically adjusts the non-Lagrangian mesh as the 
evolution proceeds (Eggleton 1971).  This recasting was adopted so that our equations had the same form as the stellar evolution 
code used to calculate the initial models, providing for a more robust ability to reconverge these 
initial models.   Along a mass loss sequence, the mass at each grid point varies. To calculate $s_i(m)$ and the 
composition, $X(m)$  at each grid point, we interpolate the initial profiles using Akima splines.  

While we focus here on donors with zero-age main sequence (MS) masses of $M_{2,0} = 10 \msun$, we calculated the 
adiabatic mass-loss response of donors with $M_{2,0}$ in the range between $1.5$--$20 \msun$ with the mass loss 
starting at evolutionary phases from the terminal age MS through the red giant and asymptotic giant branch 
phases.  The models from which we initiated mass-loss sequences we calculated using the STARS code (Eggleton 1971, 
1972, 1973; updated as described in Pols et al. 1995) for a metallicity of Z = 0.02. We reconverged each 
input initial model (to an accuracy of 
$\approx 10^{-9}$ in a norm of summed, squared relative residuals) in our own code with $M_2$ held fixed. We then 
calculated the  mass loss sequences by decrementing $M_2$ by a small amount $\Delta M_2$, based on both $M_2$ and 
the current surface entropy gradient,  and converging the model at this new $M_2$ value.  The results of the full 
study will be presented in a follow-up paper. 

Our $M_{2,0} = 10 \msun$ mass loss sequences were calculated with a numerical resolution of 599 grid points.  The 
variation of stellar radius, 
$R_2$, with $M_2$ for several of our sequences is shown 
for donors at the end of the MS, near the end of the Hertzsprung gap (HG) , and within the red giant branch 
(RGB) in Fig. 1.  The marked difference in the initial response between the RGB model's mild expansion and the 
other phase's rapid contraction is due to the transition from a radiative envelope in the less evolved donors to 
the development of a nearly flat entropy profile in the RGB stars' convective envelopes.  As the donor moves up 
the RGB, the H$\rightarrow$He burning shell thins and increases in luminosity.  This, in tandem with the resulting 
deepening of the convective envelope, produces a steepening entropy profile in the core/envelope transition region.  The steepening of  the $R_2$ response near the core/envelope boundary for more evolved models 
results from this entropy gradient.  Correspondingly, the more gradual contraction experienced during the later 
evolution of the less evolved donors results from a shallower entropy gradient throughout the donor's interior. 
The intersection of curves at $m \approx 2.3 \msun$ corresponds to the mass at the top of the H-burning shell.  

We emphasize that in the adiabatic limit, the $R_2$ response to mass loss depends only on the donor's initial 
structure through its $s$-profile.  Thus, in a CE event, the $R_2 $ vs. $M_2$ evolution is independent of other 
external factors (such as binary parameters) under the assumption that the binary companion does not affect 
the donor's structure.  This allows us to calculate in advance a library of $R_2$--$M_2$ 
tracks that form the essential input for an improved CE event outcome determination methodology. 

\begin{figure}
\plotone{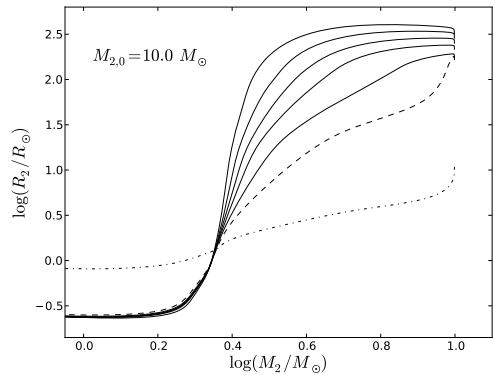}
\caption{ The radius evolution of 10 $\msun$ donors undergoing mass loss for various 
initial evolutionary states. The dot-dash and dash-line curves (with radius at the onset of the rapid mass 
loss episode, $R_{2,0} = 10.8$ and $163.3 \rsun$) illustrate the adiabatic response of donors either near the 
end of the main sequence or near the end of the HG stage respectively.  The five solid line 
curves (with $R_{2,0} = 177.0$, $216.2$, $251.4$, $287.2$, and $318.7 \rsun$) illustrate the response of 
donors at various stages on the RGB.}  
\label{fig:MvR}
\end{figure}

\section{A Self-Consistent Method for Determining CE Outcomes }

During a CE phase, mass from the donor's envelope is stripped and then ejected from the binary at the expense 
of the orbital energy.  Current prescriptions for the final outcome of this process, in terms of post-CE orbital 
separation, $a_f$, take as an \textit{a priori} input the donor's post-CE remnant mass, $M_{2,f}$.  This mass 
is typically taken to be the mass of the donor's evolved core as determined via one of several criteria applied 
to isolated stellar evolution models.  For lower MS stars on their giant branches, this approximation is reasonable 
given the steep entropy profiles across their thin-shell burning regions.  However, stars on 
the upper main sequence can have significantly shallower entropy gradients in the core/envelope boundary region.  
This leads to ambiguity in defining the core mass in these cases and produces a significant variation 
in predicted CE event outcomes (see, e.g., Tauris \& Dewi 2001).  

More importantly, in such cases, \emph{the physical motivation for correlating core mass with 
$M_{2,f}$ no longer holds}.  That is, the adiabatic mass loss sequence for lower MS donors have 
$R_2$--$M_2$ tracks that are similar to the track starting from the most-evolved donor with largest $R_{2,0}$ in 
Fig. \ref{fig:MvR}, where the steep drop-off 
in $R_2$ near the core essentially guarantees the donor contracts within its Roche lobe with $M_2$ nearly equal to the core 
mass.  For cases with more gradual $R_2$ evolution this is not guaranteed to be the case and the basic premise 
of using an \textit{a priori} determined $M_{2,f}$ becomes suspect.

To address this issue, we have developed a new method for determining CE event outcomes based not on specifying 
$M_{2,f}$, but on the $R_2$ evolution.  Specifically \emph{we assume that the CE event ends at the point the 
donor contracts within a critical radius, which we take as its Roche radius}.  To determine this point, both 
$R_2$ and the donor's Roche lobe radius, 
$R_L$, are calculated as a function of $M_2$ during the CE event.  Our mass loss $R_2$--$M_2$ tracks 
provide the first necessary input.  To determine the $R_L$--$M_2$ track during the CE event we 
assume the standard CE energy conservation (i.e., $\aCE$) prescription for determining $a_f$ 
applies along the entire course of the CE event.  That is, the orbital separation, $a$ during the 
CE phase is given by 
\begin{equation}
  -\aCE \left(\frac{G M_2 M_1}{2a} - \frac{G M_{2,i} M_{1,i}} {2a_i}\right) = E_{\mathrm{bind}}.
  \label{eq:aCE}
\end{equation}
where $G$ is Newton's gravitational constant, $\aCE$ is the CE efficiency,  and $M_1$ is the mass of the donor's 
companion. Quantities with $i$-subscripts represent quantities at the onset of the CE event.

The quantity $E_{\mathrm{bind}}$ is the difference in the donor's total energy (gravitational potential plus 
internal thermal) between its initial and current configuration.  We emphasize that this represents a subtle 
distinction relative to all previous applications of this prescription. Historically,  $E_{\mathrm{bind}}$ would 
represent the total binding energy of the envelope as determined from the \emph{initial donor model}.  There is a 
difference between the initial envelope binding energy as a function of $m$ and the donor's total binding energy 
at the corresponding total $M_2$ along the mass loss sequence.  This difference reflects the work done on overlying 
layers as the donor's mass shells expand under mass loss.  This means that $E_{\mathrm{bind}} < 0$ for a given  
$M_2$ is always greater than the initial binding energy at the corresponding $m = M_2$.  Thus, use of the initial 
model \emph{overestimates} the amount of orbital contraction produced in a CE event.  

We choose $a_i$ such that  $R_{2,i} = R_L (a_i, M_{2,i}, M_{1,i})$, using the expression from Eggleton (1983) 
to calculate $R_L$.  Modulo a few complications that arise in some cases (e.g., systems that experience delayed dynamical 
instabilities) $R_L (M_2)$ is calculated
during a CE event using $a$ as calculated from equation (\ref{eq:aCE}), assuming the mass transfer is completely 
non-conservative throughout, i.e., $M_1 = M_{1,i}$ (see Hjellming \& Taam 1991). To initiate the unstable 
mass transfer that leads to a CE event,  $d \ln R_2/ d \ln M_2 < d \ln R_L/ d \ln M_2$ initially, guaranteeing 
that $R_2 > R_L$ for some time after the onset of the CE event.  We determine the end of 
the CE event to be the point at which $R_2$ first becomes less than $R_L$.

Several points deserve to be made about this method for calculating CE outcomes.  First, the donor's 
remnant mass, $M_{2,f}$ is an output of our model, not a predetermined input.  That is, our method 
provides a self-consistent determination of $M_{2,f}$ and $a_f$.  Second, $R_L$ is a function of $M_2$, 
$M_{1,i}$ and $\aCE$, meaning that $M_{2,f}$ explicitly depends on not only the initial binary parameters, but 
also on $\aCE$.  This is in stark qualitative contrast with previous CE prescriptions.

\begin{figure}
\plotone{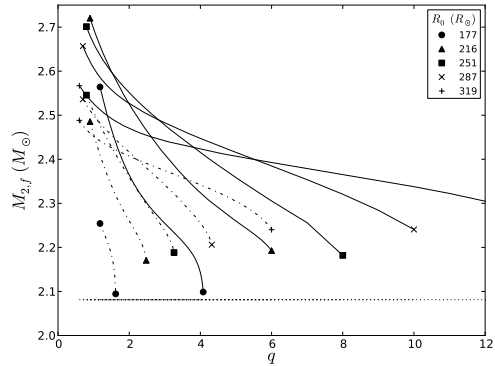}
\caption{The remnant core mass versus initial mass ratio as a function of binary and CE event
parameters.  The symbols connected by the lines denote models corresponding to a given evolutionary 
state (parameterized by $R_{2,i}$) for which a 10 $\msun$ donor star fills its Roche lobe, initiating the CE event.
The solid and dashed-dotted line illustrates the outcomes for a CE evolution with 
$\aCE=1.0$ and $\aCE= 0.5$ respectively. The dotted line represents the core mass defined by the mass 
where the hydrogen mass abundance is 0.1.}
\label{fig:M2fvq}
\end{figure}

Our method is also in contrast with the existing CE prescriptions in terms of quantitative predictions for CE 
event outcomes.  We illustrate this for the case of a $M_{2,0} = 10 \msun$ donor in Figure \ref{fig:M2fvq}. Here,
we show $M_{2,f}$ as a function of the binary's zero-age MS mass ratio, $q = M_{2,0}/M_{1,0}$ and $R_{2,i}$.  
The systematically smaller $M_{2,f}$ for smaller $\aCE$ values
result from a greater degree of orbital contraction.  For the fiducial $\aCE = 1.0$ case, $M_{2,f}$ can vary 
by as much as $\approx 20\%$ over the $q$ range producing non-merger CE events.  The development of a deep 
convective envelope and steep entropy gradient near the core/envelope boundary on the 
upper RGB produces the transition from a sharply varying remnant core mass with respect to q 
to a more gradual decrease with increasing q seen with increasing $R_{2,i}$.   

For comparison, the dotted lines in Fig. \ref{fig:M2fvq} show the outcomes if one were to use a composition 
criteria to determine the core mass. We note that this value is roughly constant from the HG 
to the RGB for a $10 \msun$ star.  
$M_{2,f}$ determined via our method can be as much as $\approx 30\%$ greater than this core mass.   

\section{Comparison with $\aCE$-Prescription Results}

To illustrate the impact of our methodology on CE event outcomes as compared with the standard 
$\aCE$-prescription we consider a population of binaries whose more massive component had an initial MS mass 
of 10 $\msun$.  Some subset of this population will undergo a CE event as this star evolves across the HG 
and up its RGB.  This initial CE event is a key step in the formation of many neutron star (NS) binary systems 
(e.g., Bhattacharya \& van den Heuvel 1991; Terman et al. 1996, 1998, Kalogera \& Webbink 1998).
For this comparison, a population of 500,000 binaries was constructed, fixing $M_{2,0} = 10 \msun$ and considering 
a set of $M_{1,0}$ that yield a uniform distribution of initial $q > 1.0$. The initial orbital separation, $a_0$, 
of this population was distributed evenly in $\ln a_0$.  For a given $(q, a_0)$ pairing, the $R_{2,i}$ at which 
the 10 $\msun$ star first filled its Roche lobe was determined.  If a CE event occurred for this system we 
determined the resulting values of $M_{2,f}$ and $a_f$ using both our self-consistent CE method 
and the standard $\aCE$-prescription, taking $\aCE = 1.0$ in both cases.

\begin{figure}
\plotone{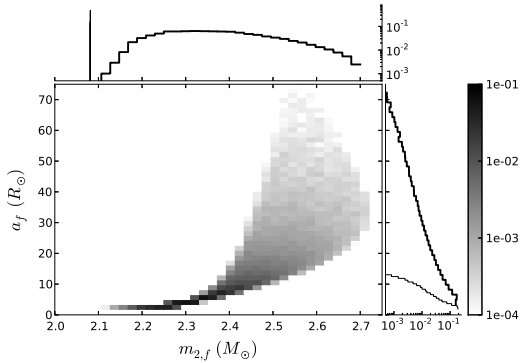}
\caption{The post-CE population of binary systems with a $10 \msun$ star that becomes the donor in the 
system's first CE event.  The grayscale shading indicates the fraction of systems 
relative to the total surviving a CE event (in this case, 26.5\% of the total number of binaries) as indicated 
by the scale on the right.  In the two side panels,  the bold lines show the corresponding histogram projected 
along each axis.  The normal-weight histograms in these two panels show the resulting distributions when the CE outcomes are 
calculated using the standard $\aCE$-prescription.} 
\label{fig:m2faf_dists}
\end{figure}

The outcome of this exercise in the form of $M_{2,f}$--$a_f$ distributions for those systems that underwent and 
survived a CE event is shown in Figure \ref{fig:m2faf_dists}.  The main panel displays the distribution of post-CE 
systems in this parameter space as calculated with our method.  This distributions peaks between 
2.25 and 2.45 $\msun$ in $M_{2,f}$ and below $\approx 10 \rsun$ in $a_f$.  However, both distributions are rather 
broad;  the $a_f$ distribution has a tail that extends to significantly larger values. The 
comparison to the $\aCE$-prescription derived distributions are displayed in the side panels, which show 
the 1-D projected distributions.  Our distribution is shown by the bold-line histogram, the $\aCE$-prescription results 
by the normal-weight lines 
with the latter scaled to the total number of systems in our distribution.  As could be easily inferred from 
Fig. \ref{fig:M2fvq}, the $M_{2,f}$ distribution in the latter case is nearly a $\delta$-function since the core 
mass of the 10 $\msun$ donor does not vary significantly over its evolution up the RGB.  The distribution in $a_f$ 
is also significantly narrower using the standard prescription as compared to our method. 

How these differences impact the subsequent evolutionary outcomes of this binary population remains to be examined.
However, we can provide some qualitative discussion here. Our method results in a 
$\approx  20\%$ higher post-CE survival rate relative to the standard prescription. This may not directly translate 
into a commensurate increase in the overall number of NS binaries produced in this channel since the core of the 
10 $\msun$ donor will begin core He-burning and continue its subsequent evolution to a core-collapse supernova. The 
kick imparted to the system during the supernova alters the orbital parameters and may unbind the system.  
Taken over a distribution of kick velocities and directions, binaries with larger orbital separations are more 
likely to be unbound during the supernova and have a wider distribution in post-supernova eccentricity (Kalogera 
1996).  Considering the subsequent evolution of those binaries that will eventually form a double NS system 
via both a second CE event and supernova (Bhattacharya \& van den Heuvel 1991), the use of our method can be expected 
to broaden the final eccentricity distribution of this population. However, our method's overall impact on the production 
rate of this (and other NS binary) population is unclear given the various competing processes. 

\section{Summary}

We have outlined a new method for calculating CE event outcomes that provides a self-consistent determination 
of post-CE core remnant masses and orbital separations in the framework of the $\aCE$ parameterization.  
This method takes full account of the donor's structural response to adiabatic mass loss based on the donor's 
internal entropy profile at the onset of the CE event.\footnote{It has come to our attention that the recent 
paper by Ge et al. (2010) also treats the adiabatic mass loss evolution.  The focus there is on the calculation 
methodology (which is similar to ours); they do include a limited discussion of the implications of their work 
that is complementary to the considerations presented here.} 
The essence of our method is the use of 
pre-calculated mass loss sequences to determine when, if ever, the donor  is able to contract within 
its Roche radius during the course of the CE event.  Here, the donor's final remnant mass depends on the 
specifics of a binary's parameters (in stark contrast to prior CE event prescriptions), in addition to assumptions 
about the CE efficiency. 

As an application of this method, we considered CE events with donors whose MS mass were $10 \msun$; this mass was 
chosen to highlight the significant impact our method has for higher mass donors that do not develop steep entropy 
gradients at their core/envelope interface until they are very far along their RGB.  For CE events initiated by such 
objects on or before their RGB tip, existing CE prescriptions adopt $M_{2,f}$ to be equal to the donor's He-core mass, 
which is essentially constant along the entire RGB.  In contrast, our method predicts an $M_{2,f}$ that varies by 
$\approx 20\%$ across the population of binaries that survive the CE event.  In addition, $a_f$ can be as large as 
$\approx 7$ times greater than the standard prescription.  Taken over a standard distribution of initial binary 
parameters, these differences result in a significantly broader distribution in both $M_{2,f}$ and $a_f$ and an overall 
CE survival rate that is $\approx 20\%$ greater than the standard prescriptions.

The vastly different distribution in post-CE parameters produced by this new methodology has implications for 
the multitude of binary populations of intermediate mass or high mass donors that experience one or more CE 
events.  Our new method could be expected to predict a broader distribution of eccentricities 
in double NS binary systems.  The overall sense of impact 
on the production rate of such systems will require detailed calculations due to several 
competing destruction processes.  Future work will involve incorporating  our methodology into 
such detailed considerations of this and the multiple other contexts in which CE outcomes play a central, if not 
uncertain, role.

\acknowledgements We wish to thank Marc van der Sluys and Tassos  Fragos for discussions and 
technical assistance.  This work was supported by NSF AST-0703950 to Northwestern University.

\end{document}